\documentclass[twocolumn,prl,showpacs,preprintnumbers,amsmath,amssymb]{revtex4-1} 
\usepackage{graphicx}
\usepackage{dcolumn}
\usepackage{bm}
\begin{document}
\title{Improved limit on a temporal variation of $m_p/m_e$ from comparisons of Yb$^+$ and Cs atomic clocks}
\author{N. Huntemann}
\author{B. Lipphardt}
\author{Chr. Tamm}
\author{V. Gerginov}
\author{S. Weyers}
\author{E. Peik}
\email{ekkehard.peik@ptb.de}
\affiliation{Physikalisch-Technische Bundesanstalt, Bundesallee 100, 38116 Braunschweig, Germany}

\date{\today}
\begin{abstract}
Accurate measurements of different transition frequencies between atomic levels of the electronic and hyperfine structure over time are used to investigate temporal variations of the fine structure constant $\alpha$ and the proton-to-electron mass ratio $\mu$. We measure the frequency of the $^2S_{1/2}\rightarrow {^2F_{7/2}}$ electric octupole (E3) transition in $^{171}$Yb$^+$ against two caesium fountain clocks as $f(E3) = 642\,121\,496\,772\,645.36(25)$~Hz with an improved fractional uncertainty of $3.9\times 10^{-16}$. This transition frequency shows a strong sensitivity to changes of $\alpha$. Together with a number of previous and recent measurements of the $^2S_{1/2}\rightarrow {^2D_{3/2}}$ electric quadrupole transition in $^{171}$Yb$^+$ and with data from other elements, a least-squares analysis yields $(1/\alpha)(d\alpha/dt)=-0.20(20)\times 10^{-16}/\mathrm{yr}$ and $(1/\mu)(d\mu/dt)=-0.5(1.6)\times 10^{-16}/\mathrm{yr}$, confirming a previous limit on $d\alpha/dt$ and providing the most stringent limit on  $d \mu/dt$ from laboratory experiments.
\end{abstract}


\maketitle

The search for variations of fundamental constants is motivated by theories unifying the fundamental interactions and is regarded as an opportunity to open a window to new physics with implications on cosmology as well as on particle physics \cite{Karshenboim2000,Uzan2003,Barrow2005,Uzan2011}. While early proposals for such a search using atomic spectroscopy have been made shortly after the discovery of the expansion of the universe \cite{Jordan1939}, sensitive observational and experimental tools became available only recently. Astrophysical observations of absorption spectra of interstellar matter have led to claims for \cite{Webb2001,Reinhold2006,Webb2011} and against \cite{Srianand2004,Levshakov2009,King2008,Thompson2009,Bagdonaite2013} variations of the fine structure constant $\alpha$ and the proton-to-electron mass ratio $\mu=m_p/m_e$ at relative uncertainties in the range $10^{-5}$ to $10^{-7}$ on a cosmological time scale of several billion years. In the laboratory, the high precision of atomic clocks that now reach relative uncertainties of $10^{-16}$ and below in frequency ratios has been used to infer limits on variations of $\alpha$ and $\mu$ in the present epoch \cite{Rosenband2008,Guena2012-b,Luo2011,Ferreira2012}.     

In this Letter we present a high-accuracy absolute frequency measurement of the $^2S_{1/2}\rightarrow {^2F_{7/2}}$ electric octupole transition in $^{171}$Yb$^+$ that possesses a strong sensitivity to changes of $\alpha$. Together with recently reported frequency measurements of the $^2S_{1/2}\rightarrow {^2D_{3/2}}$ electric quadrupole transition in the same ion \cite{Tamm2014} this allows us to constrain possible temporal changes of both transition frequencies relative to caesium clocks. Besides confirming limits on $d\alpha/dt$ in the low $10^{-17}/$yr range these data provide the most stringent limit on $d \mu/dt$ from a laboratory experiment.

\begin{figure}
\includegraphics[width=\columnwidth]{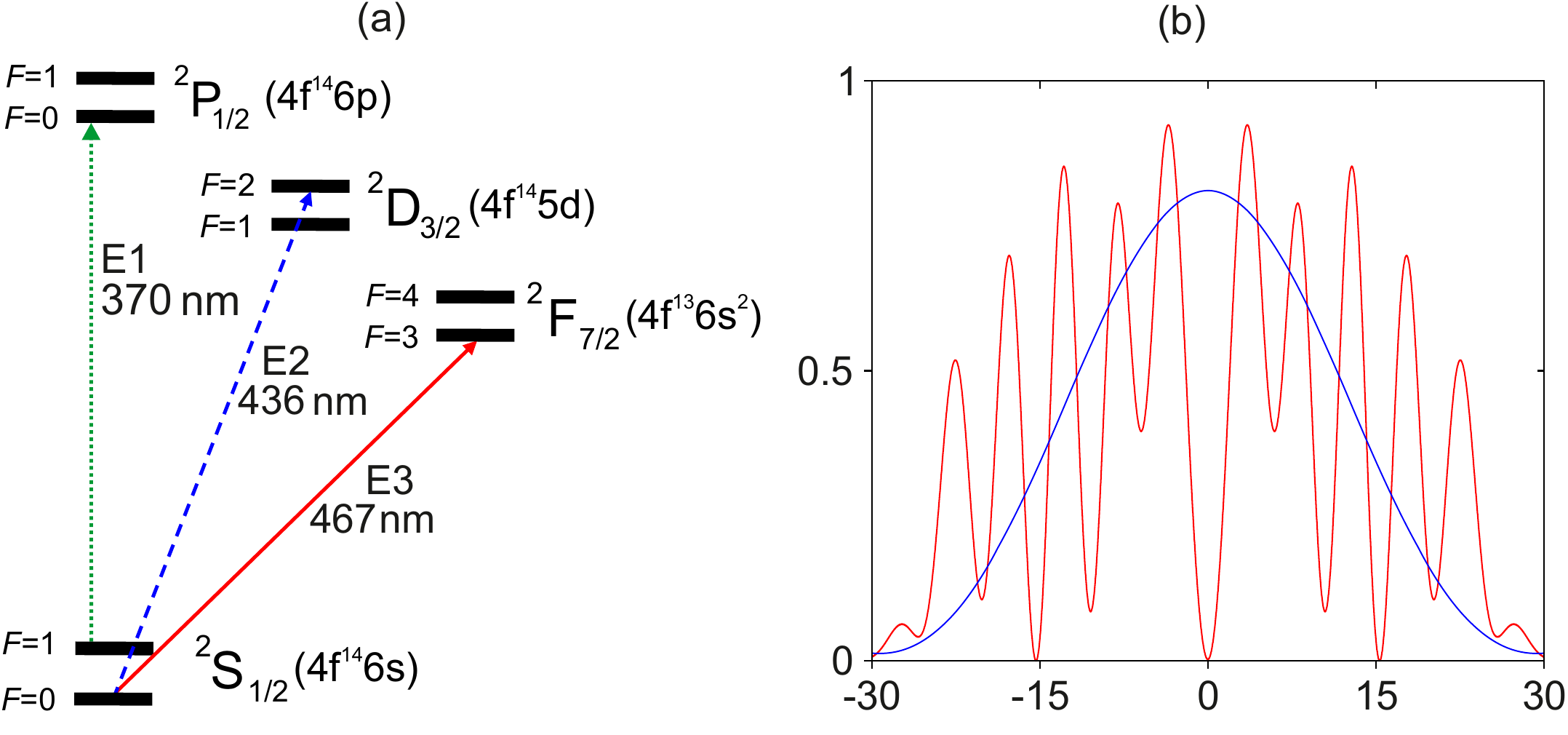}
\caption{(Color online) In (a) parts of the $^{171}$Yb$^{+}$ term scheme are shown. The strong $^2$S$_{1/2}\to {}^2$P$_{1/2}$ electric dipole (E1) transition at 370~nm is used to laser-cool the ion. The electric quadrupole (E2) and electric octupole (E3) transition serve as the reference for optical frequency standards. The respective excitation spectra are shown in (b), where the E2 transition (dashed line) is excited by 30~ms rectangular pulses and the E3 transition (solid line) using the hyper-Ramsey method with the parameters $\tau=30.5$~ms and $T=4\tau$ (see text). \label{LevelScheme}}
\end{figure}

$^{171}$Yb$^+$ is particularly attractive for a search for variations of fundamental constants because there are two transitions with low natural linewidth from the ground state to metastable states that have rather different electronic configurations [see Fig.~\ref{LevelScheme}(a)]. In case of the $^2$S$_{1/2} (F=0)\to {}^2$D$_{3/2} (F=2,m_F=0)$ electric quadrupole (E2) transition at 436~nm  the $6s$ valence electron is promoted to the $5d$ level, while on the $^2$S$_{1/2} (F=0)\to {}^2$F$_{7/2} (F=3,m_F=0)$  electric octupole (E3) transition at 467~nm an electron is taken from the fully occupied $4f$ shell to fill the $6s$ shell. This can also be seen as the excitation of a hole state from the $6s$ to the $4f$ shell. Consequently, variations in $\alpha$ would lead to opposite shifts of the frequencies of the E2 and E3 transitions. As a result of the large proton number of Yb$^+$, the relevant level energies contain important relativistic contributions and are therefore particularly sensitive to variations of $\alpha$ \cite{Lea2007,Flambaum2009}. 

To realize optical frequency standards based on the E2 and the E3 transition  of $^{171}$Yb$^+$, a single laser-cooled ion is confined in a cylindrically symmetric radio frequency Paul trap \cite{Tamm2014,Huntemann2012}. Apart from the laser systems probing the E2 and E3 transitions, the same experimental setup is utilized for both standards.  Laser cooling is performed on the strong $^2$S$_{1/2}\to {}^2$P$_{1/2}$ electric dipole transition at 370~nm while a repump laser at 935~nm prevents population trapping in the $^2$D$_{3/2} (F=1)$ state due to spontaneous decay from the $^2$P$_{1/2}$ level. Each measurement cycle starts with a 15~ms period of laser cooling followed by optical pumping to the $^2$S$_{1/2} (F=0)$ ground state. The stronger magnetic field applied during laser cooling is reduced to $3.58(1)~\mu$T at one of three orientations that are mutually orthogonal with an uncertainty of $\leq$1$^\circ$. This enables an efficient cancellation of tensorial shifts \cite{Itano2000}. Under these conditions, the reference transition is interrogated by the probe laser, while mechanical shutters block the cooling laser and the repump laser beams. Successful excitation to the excited state is indicated by absence of fluorescence at the beginning of the subsequent cooling period. The E2 transition is probed by single 30~ms long rectangular probe laser pulses and the resulting spectrum is depicted in Fig.~\ref{LevelScheme}(b)(dashed line). Due to the natural lifetime of 52.7(2.4)~ms of the $^2$D$_{3/2}$ level \cite{Yu2000}, the maximum resonant excitation probability is limited to about 80\%. The realization of a frequency standard based on the E2 transition with a fractional systematic uncertainty of $1.1\times 10^{-16}$ and a measurement of the transition frequency versus caesium fountain clocks with a total relative uncertainty of $5.2\times 10^{-16}$ are discussed in detail in Ref.~\cite{Tamm2014}. 

The extraordinary long natural lifetime of the $^2$F$_{7/2}$ state on the order of years \cite{Roberts2000} permits an observation of the transition with a linewidth essentially determined by the frequency instability of the probe laser system. Because of the extremely small oscillator strength of the octupole transition, its excitation requires particularly high spectral power density. The required intensity leads to nonresonant couplings to higher-lying levels and thereby introduces a significant light shift of the transition frequency. In a first realization of an optical frequency standard based on the E3 transition, real-time extrapolation to zero probe laser intensity was used to cancel this shift \cite{Huntemann2012}. The achieved relative uncertainty due to the light shift of $0.42\times 10^{-16}$ can be significantly reduced by application of a generalized Ramsey scheme \cite{Yudin2010,Huntemann2012b}. Here, the effect of the light shift is compensated by a step of the probe laser frequency that approximates the light shift during the interaction periods and a third echo-type pulse between the Ramsey pulses suppresses the linear dependence of the position of the central fringe on the light shift. The calculated lineshape of the excitation spectrum for a   Ramsey pulse duration of $\tau=30.5$~ms and a free evolution period $T=4\tau$ is shown in Fig.~\ref{LevelScheme}(b) (solid line). As proposed in Ref.~\cite{Huntemann2012b}, the efficient suppression of the light shift can be ensured by interleaved measurements with single-pulse Rabi spectroscopy. With this technique and by averaging the realized frequency for three orthogonal orientations of the magnetic field to cancel tensorial shift effects, the previously reported systematic uncertainty of $0.71\times 10^{-16}$ \cite{Huntemann2012} has been reduced to $0.50\times 10^{-16}$. This total systematic uncertainty is essentially determined by the uncertainty of the quadratic Stark shift induced by thermal radiation at room temperature. 

The frequency of the unperturbed $^2$S$_{1/2} (F=0)\to {}^2$F$_{7/2} (F=3,m_F=0)$ transition has been measured versus the two caesium fountain clocks CSF1 and CSF2 of our laboratory \cite{Weyers2001,Gerginov2010} by means of a fiber laser based frequency comb \cite{Lipphardt2009}. To enhance the frequency stability of CSF2, a microwave dielectric resonator oscillator serving as the local oscillator for CSF2 was stabilized using the E3 probe laser system \cite{Lipphardt2009,Weyers2009,Tamm2014}. The total measurement time of 350\ 000~s results in relative statistical uncertainties of $2.5\times 10^{-16}$ and $2.0\times 10^{-16}$ for the measurements with CSF1 and CSF2. The systematic uncertainties of the caesium fountains are $7.3\times 10^{-16}$ for CSF1 and $4.0\times 10^{-16}$ for CSF2. Averaging the two measured frequencies and combining the respective uncertainties with the $0.50\times 10^{-16}$ uncertainty of the optical standard yields the frequency of the unperturbed octupole transition as $f(E3) = 642\,121\,496\,772\,645.36(25)$~Hz. The individual results of $f(E3) - 0.17(50)$~Hz and $f(E3)+0.07(29)$~Hz of CSF1 and CSF2 are in very good agreement. The new measurement of $f(E3)$ reduces the uncertainty by more than a factor of two in comparison to our previously reported value \cite{Huntemann2012}, constituting one of the most precise measurements of optical transition frequencies \cite{Falke2014,Letargat2013}.

For the analysis of the clock comparisons in terms of variations of dimensionless fundamental constants, we use the following parametrization of the atomic transition frequencies \cite{Karshenboim2000}: The optical transition frequency between levels of the electronic structure can be written as 
\begin{equation}
f=cR_{\infty} C F(\alpha),
\end{equation} 
and similarly the hyperfine structure transition frequency as
\begin{equation}
f_{H}=\alpha^2 cR_{\infty} C_H F_H(\alpha)G(\mu_N/\mu_B).
\end{equation} 
The Rydberg frequency $cR_{\infty}=m_ee^4/(8\epsilon_0^2h^3)$ gives the non-relativistic energy scaling of electronic transitions in atoms and molecules and therefore cancels in frequency ratio measurements of atomic clocks. The numerical factors $C$ and $C_H$ describe the non-relativistic atomic structure. They depend on the quantum numbers characterizing the state and are assumed to be constant. The dimensionless functions $F(\alpha)$ and $F_H(\alpha)$ describe relativistic level shifts and the function $G(\mu_N/\mu_B)$ contains the dependence on nuclear structure via the nuclear magnetic moment. The dependence on the proton-to-electron mass ratio enters here via $\mu_N/\mu_B\propto 1/\mu$, together with a dependence on the 
strong-interaction parameter $X_q$, that denotes the ratio between the average quark mass and the quantum chromodynamic scale $\Lambda_{QCD}$ \cite{Flambaum2006,Dinh2009}. It can be seen that in this parametrization any ratio of electronic transition frequencies obtained from optical  standards is sensitive to a variation of $\alpha$ only. Ratios of electronic to hyperfine  transition frequencies are sensitive to variations of $\alpha,~\mu$ and $X_q$. 
So-called absolute frequencies of optical frequency standards measured with reference to caesium clocks and expressed in the unit Hz of the International System of Units (SI) can easily be converted into this type of ratio by dividing the frequency by the conventional value of the $^{133}$Cs ground state hyperfine transition frequency of $9\,192\,631\,770$~Hz. 
Allowing for variations of  $\alpha,~\mu$ and $X_q$, the relative change of a frequency ratio $R=f/f_{H}$ is given by:
\begin{equation}
\frac{1}{R}\frac{dR}{dt}=(K-K_{H}-2) \frac{1}{\alpha}\frac{d\alpha}{dt}+ \frac{1}{\mu}\frac{d\mu}{dt} - \kappa \frac{1}{X_q}\frac{dX_q}{dt},
\end{equation} 
where the sensitivity factors $K$ and $\kappa$ are determined by 
\begin{equation}
K=\frac{1}{F}\frac{dF}{d\alpha}
\end{equation} 
and 
\begin{equation}
\kappa=\frac{1}{G}\frac{dG}{dX_q}
\end{equation} 
and can be obtained from atomic and nuclear structure calculations \cite{Flambaum2009,Flambaum2006,Dinh2009}. A search for variations of $\alpha$ and $\mu$ with caesium clocks benefits from the finding that the sensitivity of the $^{133}$Cs ground state hyperfine frequency to variations of the quark masses is much smaller than to variations of $\alpha$  and $\mu$: For $^{133}$Cs, $\kappa(Cs)=0.002$ \cite{Dinh2009} and 
$K_{H}(Cs)=0.83$ \cite{Flambaum2006}.

\begin{figure}
\includegraphics[width=.8\columnwidth]{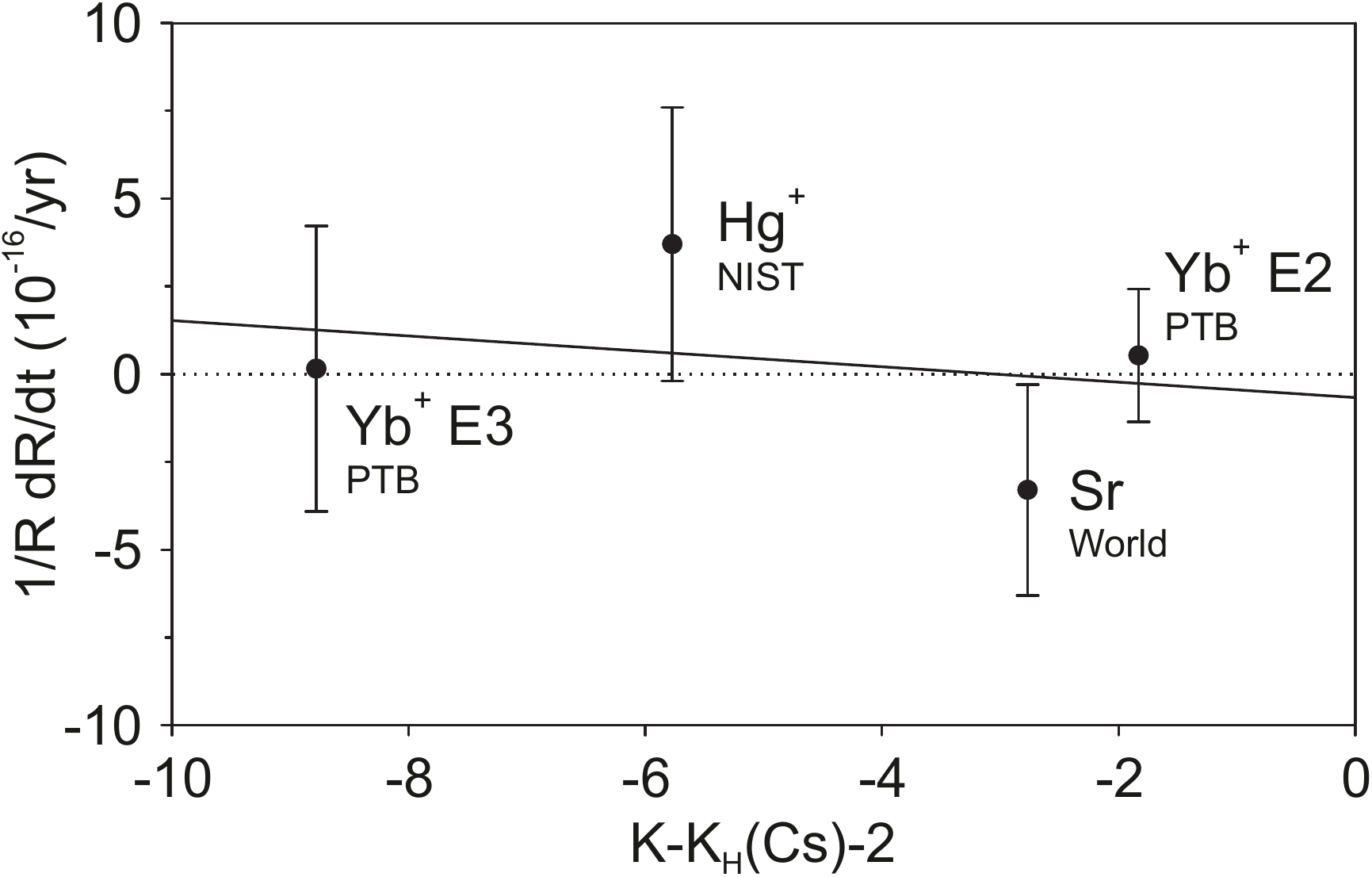}
\caption{Relative temporal changes of ratios $R$ between optical transition frequencies and the  $^{133}$Cs ground state hyperfine transition frequency, versus the sensitivity of the respective combination of transitions to changes of $\alpha$ (see Tab.~1). The solid line is the result of a weighted linear regression. A nonzero slope of this line would indicate a variation of $\alpha$, while the intercept is predominantly determined by a variation of $\mu$. \label{DriftRate}}
\end{figure}

Table 1 summarizes the limits on temporal variations of absolute frequencies for the four optical transitions for which the most precise data is available, including the two $^{171}$Yb$^+$ transitions investigated in this experiment. Using this data a linear regression of $(1/R)dR/dt$ as a function of $K-K_H(Cs)-2$ can be performed, yielding $(1/\alpha)  d\alpha/dt$ as the slope (see Fig.~\ref{DriftRate}). The value of $(1/\mu)  d\mu/dt$ can be obtained from the intercept, after subtracting a small contribution for a possible variation of the quark masses, for which we use the result
\begin{equation}
\kappa(Cs)\frac{1}{X_q}\frac{dX_q}{dt}=0.14(9)\times 10^{-16}/{\mathrm yr}
\end{equation} 
that has been inferred from a comparison of the $^{87}$Rb and $^{133}$Cs hyperfine frequencies \cite{Guena2012-b}. 

\begin{table*}
\caption{Experimental limits on temporal variations of optical atomic
transition frequencies relative to Cs clocks. The sensitivity factors $K$ to changes of $\alpha$ are taken from Ref. \cite{Flambaum2009}.}
\newcommand{\m}{\hphantom{$-$}}
\newcommand{\cc}[1]{\multicolumn{1}{c}{#1}}
\renewcommand{\tabcolsep}{2pc} 
\renewcommand{\arraystretch}{1.2} 
\begin{tabular}{@{}lrrr}
\hline
Atom, transition  &  $K$ & $(1/R)dR/dt$ ($10^{-16}$/yr)&  Reference  \\
\hline
$^{87}$Sr, $^1S_0\rightarrow {^3P_0}$ & 0.062& $-3.3\pm 3.0$ & \cite{Letargat2013}\\
$^{171}$Yb$^+$, $^2S_{1/2}\rightarrow {^2D_{3/2}}$ & 1.0 & $0.5\pm 1.9$ 
&   \cite{Tamm2014}\\
$^{171}$Yb$^+$, $^2S_{1/2}\rightarrow {^2F_{7/2}}$  & -6.0 & $0.2 \pm 4.1$
&   this work\\
$^{199}$Hg$^+$, $^2S_{1/2}\rightarrow {^2D_{5/2}}$ & -2.9& $3.7 \pm 3.9$  & \cite{Fortier2007}\\
\hline
\end{tabular}
\end{table*}

The result for the data in table 1 is:
\begin{equation}
\frac{1}{\alpha}\frac{d\alpha}{dt}=-0.22(59)\times 10^{-16}/{\mathrm yr}
\end{equation} 
\begin{equation}
\frac{1}{\mu}\frac{d\mu}{dt}=-0.5(2.4)\times 10^{-16}/{\mathrm yr}
\end{equation} 
and is consistent with the constancy of these constants.

\begin{figure}
\includegraphics[width=.9\columnwidth]{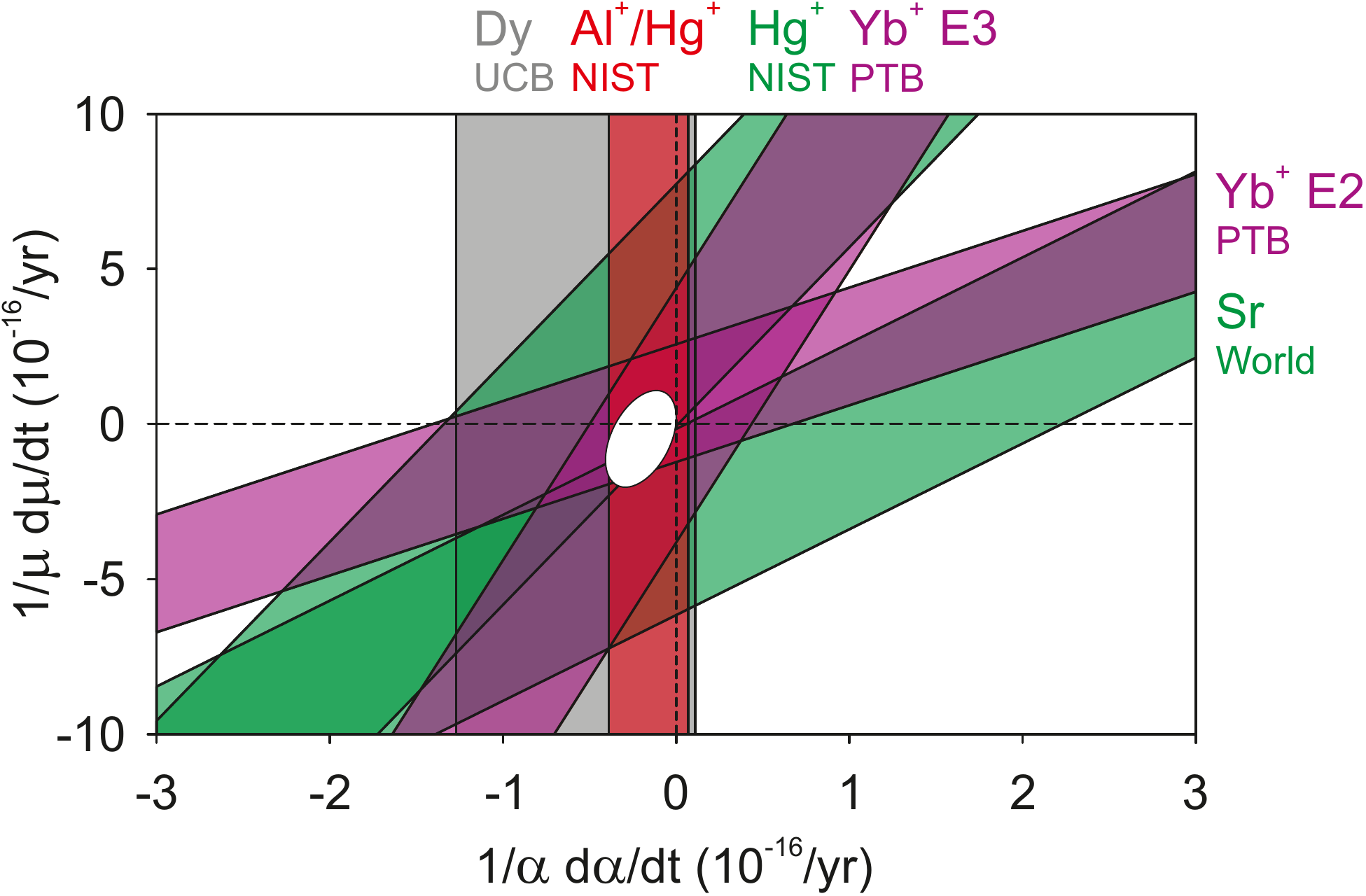}
\caption{(Color online) Constraints on temporal variations of $\alpha$ and $\mu$ from comparisons of atomic transition frequencies from Refs.~\cite{Fortier2007,Rosenband2008,Leefer2013,Letargat2013,Tamm2014} and this work. Filled stripes mark the 
$1\sigma$-uncertainty regions of individual measurements and the central blank region is bounded by the standard uncertainty ellipse resulting from the combination of all data. \label{ellipse}}
\end{figure}

Combining these results with other stringent limits on $(1/\alpha)  d\alpha/dt$ obtained from comparisons of transition frequencies in Al$^+$ and Hg$^+$ \cite{Rosenband2008} and in Dy \cite{Leefer2013} constrains  $(1/\mu)  d\mu/dt$ further. Fig.~\ref{ellipse} shows limits obtained from individual experiments as stripes marking the $1\sigma$-uncertainty ranges in the $( d\alpha/dt, d\mu/dt)$ plane. The small contribution from ${dX_q}/{dt}$ is contained in the positions and widths of the stripes for the Yb$^+$, Hg$^+$ and Sr absolute frequency measurements. Also shown is the standard uncertainty ellipse \cite{Beringer2012}, calculated as the contour with normalized quadratic deviation $\chi^2=1 + \chi^2_{min}$ where  $\chi^2_{min}=2.38$ is the minimum found in the least-squares fit. Taking the projections of the ellipse on the coordinate axes as the uncertainty ranges, one obtains:
\begin{equation}
\frac{1}{\alpha}\frac{d\alpha}{dt}=-0.20(20)\times 10^{-16}/{\mathrm yr}
\end{equation} 
\begin{equation}
\frac{1}{\mu}\frac{d\mu}{dt}=-0.5(1.6)\times 10^{-16}/{\mathrm yr}.
\end{equation} 
The limit on changes of $\mu$ is about two times more stringent than from the most comprehensive previous analysis that was done without the recent data from Yb$^+$ \cite{Guena2012-b}. 

Future progress in the uncertainty of $d\alpha / dt$ can be expected from direct measurements of optical frequency ratios because lower systematic uncertainties than in measurements with caesium clocks can be obtained. From the measurements reported here and in \cite{Tamm2014} one obtains for the ratio of the two Yb$^+$ reference frequencies $f(E3)/f(E2)=0.932\, 829\, 404\,530\,966\,29(55)$. The relative uncertainty of $5.9\times10^{-16}$ is dominated by the systematic uncertainty of the caesium fountain clocks of $5.4\times10^{-16}$. The latter represents an upper limit because some contributions to the systematic shift can be assumed to influence both measurements in the same way. 

This work was supported by Deutsche Forschungsgemeinschaft in QUEST and by the European Metrology Research Programme (EMRP) in project SIB04. The EMRP is jointly funded by the EMRP participating countries within EURAMET and the European Union. 

During the preparation of this manuscript we became aware of a related analysis, using part of the data on Yb$^+$ presented here together with independent measurements of Yb$^+$ frequencies at NPL \cite{Godun2014}.

\end{document}